\documentclass[twoside]{article}
\usepackage{fleqn}
\usepackage{espcrc2}
\usepackage{acromake}
\usepackage{axodraw}
\makeatletter
\def\@oddhead{\ifnum\c@page=1\reset@font\hfil\rm EPTCO-98-003\else\@empty\fi}
\let\@evenhead\@empty
\def\@oddfoot{\reset@font\hfil\rm\thepage\hfil}
\let\@evenfoot\@oddfoot
\makeatother
\newcommand{\mA}{m_{\scriptscriptstyle A}}
\newcommand{\mB}{m_{\scriptscriptstyle B}}
\newcommand{\LO}{\Lambda^0}
\newcommand{\SO}{\Sigma^0}
\newcommand{\SM}{\Sigma^-}
\newcommand{\SP}{\Sigma^+}

\newcommand{\XO}{\Xi^0}
\newcommand{\XM}{\Xi^-}
\newcommand{\vev}[1]{\left\langle#1\right\rangle}
\renewcommand{\ACRcntb}{0}
\acromake{HSD}{HSD}{hyperon semi-leptonic decay}
\acromake{CoM}{CoM}{centre-of-mass}
\title{SU(3) Breaking in Hyperon Beta Decays:
  a Prediction for $\XO\to\SP{e}\bar\nu$
  \thanks{Presented at the III Int.\ Conf.\ on Hyperons, Beauty and
    Charm Hadrons, (Genova, June-July 1998).}%
}

\author{Philip G. Ratcliffe
  \address{%
    II Facolt\`a di Scienze,
    Universit\`a di Milano---sede di Como,
    via Lucini 3, 22100 Como, Italy \\
    and Istituto Nazionale di Fisica Nucleare---sezione di Milano}%
  \thanks{\tt pgr@fis.unico.it}%
}
\begin{document}

\begin{abstract}
  SU(3) breaking in hyperon semi-leptonic decays is discussed. The
  SU(3) parameters $F$ and $D$, relevant to the ``proton-spin
  puzzle'', are extracted and a prediction is presented for the
  decay $\XO\to\SP{e}\bar\nu$, currently under study by the KTeV
  collaboration. The values found are $g_1/f_1=1.16\pm0.03\pm0.01$
  and $\Gamma=(0.80\pm0.03\pm0.01)\cdot10^6\,\mbox{s}^{-1}$.
\end{abstract}

\maketitle

\section{Introduction}

Beside the accurately measured $\beta$-decay rate and angular
asymmetries for the neutron, there is also a solid body of data
regarding the rest of the baryon octet \cite{PDG96a}. In SU(3) such
decays are described via two parameters, $F$ and $D$, relating to
strong-interaction effects and two further parameters, $V_{ud}$ and
$V_{us}$, the CKM matrix elements (possible contributions of heavier
flavours may be safely neglected). The $F$ and $D$ parameters are
important, as they appear in the well-known Ellis-Jaffe sum rule
\cite{Ell74a}; a 15\% reduction in the ratio $F/D$ from its accepted
value ($\sim0.6$) would remove the discrepancy with polarised DIS data
and alleviate the ``proton-spin puzzle'' \cite{Clo93a}.

As SU(3) is generally violated at the 10\% level, it is important to
develop a reliable description of the breaking. A serious test of any
scheme proposed lies in the predictions made for new decays. The process
$\XO\to\SP{e}\bar\nu$, soon to be measured accurately by the KTeV
collaboration at Fermilab \cite{Monnier:1998a}, will provide just such a
test. In this talk, following an outline of the data and a scheme to
describe SU(3) breaking, I shall present a prediction for the above
decay \cite{Ratcliffe:1998su}.

\section{The HSD Data}

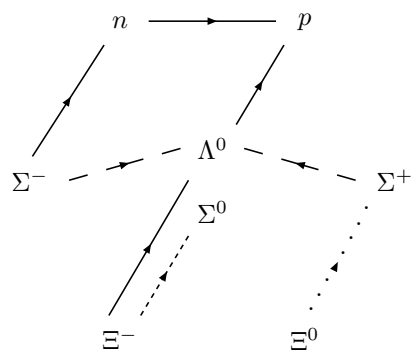
\begin{figure}[hbt]
\begin{center}
\begin{picture}(140,140)(-70,-50)
\SetWidth{1.0}
\SetScale{0.6}
\ArrowLine    (-113,  15)( -68,  85)      
\ArrowLine    (  15,  35)(  45,  85)      
\ArrowLine    ( -65, -85)( -15,   0)      
\ArrowLine    ( -40, 100)(  40, 100)      
\DashArrowLine( -90,   0)( -20,  20){15}  
\DashArrowLine(  90,   0)(  20,  20){15}  
\DashArrowLine( -45, -85)( -15, -35){3}   
\put          (  36, -49){.}
\put          (  39, -44){.}
\put          (  42, -38){.}
\ArrowLine    (  77, -55)(  78, -53)
\put          (  48, -28){.}
\put          (  51, -22){.}
\put          (  54, -17){.}
\put          (  57, -11){.}
  \Text( -35,  60)[c]{$n$}
  \Text(  35,  60)[c]{$p$}
  \Text( -69,   0)[c]{$\Sigma^-$}
  \Text(   0,  12)[c]{$\Lambda^0$}
  \Text(   0, -12)[c]{$\Sigma^0$}
  \Text(  69,   0)[c]{$\Sigma^+$}
  \Text( -35, -60)[c]{$\Xi^-$}
  \Text(  35, -60)[c]{$\Xi^0$}
\end{picture}
\caption{\label{SU3scheme}%
  The SU(3) scheme of the measured baryon octet $\beta$-decays: the
  solid lines represent decays for which both rates and asymmetry
  measurements are available; the long dash, only rates; the short dash,
  $f_1=0$ decays; and the dotted line, the forthcoming KTeV data.}
\end{center}
\end{figure}

Fig.~\ref{SU3scheme} displays the measured baryon octet $\beta$-decays,
indicating the nature of the data available.
Note that a few of the decays have also been studied in the $\mu$
mode. The present world data on \HSD are collected in
table~\ref{HSDdata}.
\begin{table*}[hbt]
  \def\arraystretch{1.1}
  \doublerulesep=0.2pt
  \setlength{\tabcolsep}{1.5pc}
  \catcode`!=\active \def!{\hphantom{0}}
  \catcode`?=\active \def?{\hphantom{-}}
  \def\F#1{\hbox to0pt{$\;^{#1}$\hss}}
\caption{\label{HSDdata}%
  The present world \protect\HSD rate and angular-correlation data
  \protect\cite{PDG96a}. The numerical values marked $g_1/f_1$ are
  those extracted from angular correlations.}
\begin{tabular*}{\textwidth}{r@{$\,\to\,$}l@{\extracolsep{\fill}}cccl}
  \hline\hline
  \multicolumn{2}{c}{Decay} &
  \multicolumn{2}{c}{Rate ($10^6$\,s$^{-1}$)} & $g_1/f_1$ & $g_1/f_1$ \\
  \cline{3-4}
  $A$   & $B\ell\nu\quad$ & $\ell=e^\pm$ & $\ell=\mu^-$
  & $\ell=e^-$  & SU(3) \\
  \hline
  $n$   & $p$   & $1.1274\pm0.0025\F{a}$ &
  & $?1.2601\pm0.0025$ & $F+D$ \\
  $\LO$ & $p$   & $3.161!\pm0.058!$      & $0.60\pm0.13$
  & $?0.718!\pm0.015!$ & $F+D/3$ \\
  $\SM$ & $n$   & $6.88!!\pm0.23!!$      & $3.04\pm0.27$
  & $-0.340!\pm0.017!$ & $F-D$ \\
  $\SM$ & $\LO$ & $0.387!\pm0.018!$      &
  & & $-\sqrt{2/3}\,D$\F{b} \\
  $\SP$ & $\LO$ & $0.250!\pm0.063!$      &
  & & $-\sqrt{2/3}\,D$\F{b} \\
  $\XM$ & $\LO$ & $3.35!!\pm0.37\F{c}!!$ & $2.1!\pm2.1\F{d}!$
  & $?0.25!!\pm0.05!!$ & $F-D/3$ \\
  $\XM$ & $\SO$ & $0.53!!\pm0.10!!$      &
  & & $F+D$ \\
  \hline\hline
\end{tabular*}
{\small
  $^a$~Rate given in units of $10^{-3}\,$s$^{-1}$.
  $^b$~Absolute expression for $g_1$ given ($f_1=0$).
  $^c$~Scale factor 2 included in error (PDG practice for discrepant data).
  $^d$~Data not used in these fits.}
\end{table*}
Let me remark that several of the rates and asymmetries have now been
measured to better than 5\%.

Although not evident from the table, there is a discrepancy in the
neutron $\beta$-decay data. However, the precision there (around
0.2\%) is far beyond the needs of the present analysis. To avoid
clouding the SU(3) breaking interpretation of the full data set, I
follow the PDG practice and rescale the discrepant data by their
resulting $\chi^2$ (see \cite{Ratcliffe:1996fk}): the errors in both
$\Gamma$ and $g_1/f_1$ for the neutron become 0.0055.  This removes
any anomalous contribution to the octet decay $\chi^2$, thus allowing
a fair comparison of fits.

\section{SU(3) Breaking and Fit Results}

SU(3) breaking can be well described using \CoM, or recoil, corrections
\cite{Ratcliffe:1996fk,Don87a,Ratcliffe:1990dh}. One approach, $A$ here,
is to account for the extended nature of the baryon by applying momentum
smearing to its wave function. Thus, \CoM corrections to $g_1$ for the
decay $A{\to}B\ell\nu$ lead to
\[
  g_1
  =
  g_1^{\mathrm{SU(3)}}
    \left[ 1 - \frac{\vev{p^2}}{3\mA\mB}
      \left( \frac14 + \frac{3\mB}{8\mA} + \frac{3\mA}{8\mB} \right)
    \right].
\]

Approach $B$ is rather similar: the breaking is related to
mass-splitting effects in the interaction Hamiltonian via first-order
perturbation theory \cite{Ratcliffe:1997ys}. The correction turns out
equivalent to a linearised version of the above:
\[
  g_1
  = g_1^{\mathrm{SU(3)}}\,\left[1-\epsilon(\mA+\mB)\strut\right].
\]

Note that in both approaches the corrections are normalised to the
reference-point correction for $g_1^{n{\to}p}$ and depend on just one
new parameter ($\vev{p^2}$ and $\epsilon$). Corrections to $f_1$ are
found to be negligible in $A$ and are assumed so in $B$, in accordance
with the Ademollo-Gatto theorem. A further global $\sim$2\%
normalisation correction to the $|\Delta{S}{=}1|$ rates marginally
improves the fit without altering the results; in \cite{Don87a} a larger
value, $\sim$8\%, was used; however, this worsens present-day fits.

Table~\ref{table:results} displays the results of three fits: $S$
(symmetric), $A$ and $B$.
\begin{table}[bt]
  \def\arraystretch{1.1}
  \doublerulesep=0.2pt
  \setlength{\tabcolsep}{6pt}
\caption{\label{table:results}%
  SU(3) symmetric and breaking fits, including $V_{ud}$
  and $V_{us}$ from nuclear \emph{ft} and $K_{\ell3}$ analyses.}
\begin{tabular}{@{\extracolsep{\fill}}ccccc}
  \hline\hline \raisebox{1.0ex}{\strut}
  Fit & $V_{ud}$  & $F$ & $D$ & 
  \raisebox{0.7ex}{$\chi^2\!\!$}/\raisebox{-0.3ex}{DoF} \\
  \hline
  $S$ & 0.9749(3) & 0.465(6) & 0.798(6) & 2.3 \\
  $A$ & 0.9743(4) & 0.460(6) & 0.806(6) & 1.2 \\
  $B$ & 0.9744(4) & 0.459(6) & 0.807(6) & 1.2 \\
  \hline\hline
\end{tabular}
\end{table}
With regard to these fits, a few clarifying remarks are useful. The value
of $V_{ud}$ (and hence $V_{us}$, fixed here via CKM unitarity) is mainly
determined by the super-allowed nuclear \emph{ft} values and so-called
$K_{\ell3}$ analyses. However, when $V_{ud}$ and $V_{us}$ are extracted
from \HSD data alone, all parameter values remain very similar. Indeed,
$F$ and $D$ are quite insensitive to the breaking schemes used.

\section{A Prediction}

Table~\ref{table:predictions} compares the predictions obtained for
$\XO\to\SP{e}\bar\nu$ from the above three fits. Recall that
$g_1/f_1=F+D$ for this decay, thus allowing for important cross
checks.
\begin{table}[t]
  \def\arraystretch{1.1}
  \doublerulesep=0.2pt
  \def\F#1{\hbox to0pt{$\;^{#1}$\hss}}
  \setlength{\tabcolsep}{10pt}
\caption{\label{table:predictions}%
  The axial coupling, rate and branching fraction ($B$) for
  $\XO\to\SP{e}\bar\nu$. The errors are those returned by the fitting
  routine.}
\begin{tabular}{cccc}
  \hline\hline
  Fit & $g_1/f_1$ & $\Gamma$ ($10^6\,$s$^{-1}$) & $B$ ($10^{-4}$) \\
  \hline
  $S$ & 1.26(0)\F{a} & 0.89(1) & 2.58(05) \\
  $A$ & 1.17(3)      & 0.80(3) & 2.32(10) \\
  $B$ & 1.14(3)      & 0.78(3) & 2.26(12) \\
  \hline\hline
\end{tabular}
{\small
  $^a$~Zero error is assigned to $g_1/f_1$ in the symmetric fit as it
  would be that of neutron $\beta$-decay.}
\end{table}
The variation between the two SU(3) breaking fits lies within the
statistical errors, I therefore combine them both and obtain the
following mean values:
\[
  g_1/f_1 = 1.16\pm0.03\pm0.01,
\]
\[
  \Gamma = (0.80\pm0.03\pm0.01) \cdot 10^6\,\mbox{s}^{-1},
\]
where the second error is an estimate of the systematic uncertainty
due precisely to the differences between the two fits.

By way of comparison, let me now very briefly examine a $1/N_c$ approach
\cite{Flores-Mendieta:1998ii}: the fit there resulted in $F/D=0.46$ and
for $\XO\to\SP{e}\bar\nu$ gave
\[
  f_1=1.12
  \quad\mbox{and}\quad
  g_1=1.02
  \quad\mbox{(fit $B$ of ref.~\cite{Flores-Mendieta:1998ii}),}
\]
or $g_1/f_1=0.91$ and $\Gamma=0.68\cdot10^6\,$s$^{-1}$. Note that both
$g_1/f_1$ and $\Gamma$ are much smaller than those presented here, which
are in turn much smaller than the na{\"\i}ve SU(3) fit. The various
possibilities should be distinguishable in an experiment with good
statistics, such as KTeV \cite{Monnier:1998a}.

To understand the difference, recall that the analysis of
ref.~\cite{Flores-Mendieta:1998ii} includes baryon-decuplet non-leptonic
decay data, which dominate; and the overall fit, in terms of $\chi^2$,
is poor. However, when applied to the \HSD data alone, the approach
produces results similar to those reported here \cite{Manohar:pc}.

\section{Conclusions}

Before concluding, let me call attention to a point that is all too
often overlooked: although easier to analyse, the data based on angular
correlations alone present absolutely \emph{no} evidence for SU(3)
breaking. Furthermore, they severely lack in statistical power.
\emph{Only full analyses can be expected to display the true picture}
\cite{Ratcliffe:1996fk}.

A full comprehension of SU(3) violation is still wanting: witness the
octet-decuplet discrepancy and the $|\Delta{S}{=}1|$ uncertainties noted
above; moreover, the system is not yet truly over-constrained. In this
context, I might also mention another decay (already measured but not
accurately so) for which large corrections are expected: namely,
$\XM\to\SO{e}\bar\nu$. There too $g_1/f_1=F+D$, allowing for additional
sensitive cross checks.

Concluding then, I would stress that while the data do clearly
manifest significant departures from SU(3), the mass-splitting driven
schemes discussed here provide a perfectly adequate description. That
said, there is clearly still much to be understood: \emph{e.g.}, the
long-standing question of second-class currents. Thus, any new
\emph{precise} data are more than welcome and the contribution of the
KTeV collaboration will be invaluable.

\section{Acknowledgments}

The author happily thanks Prof.~E.C. Swallow for much helpful
information and comment.


\end{document}